\documentclass[12pt]{iopart}
\usepackage{iopams}  

\usepackage[breaklinks,colorlinks,citecolor=blue]{hyperref}

\usepackage{graphicx}

\usepackage[utf8]{inputenc}

\usepackage{amssymb}

\newcommand{\be}{\begin{equation}}
\newcommand{\ee}{\end{equation}}
\newcommand{\bea}{\begin{eqnarray}}
\newcommand{\eea}{\end{eqnarray}}
\newcommand{\Dd}{\mathrm{d}}

\begin{document}

\title[Teaching cosmology with special relativity]{Teaching cosmology with special relativity: Piecewise inertial frames as a model for cosmic expansion}
\author{Markus P\"ossel}

\address{Haus der Astronomie and Max Planck Institute for Astronomy, K\"onigstuhl 17, 69124 Heidelberg, Germany}
\ead{poessel@hda-hd.de}
\vspace{10pt}
\begin{indented}
\item[]\today
\end{indented}

\begin{abstract}
This article presents a simple model that reproduces key concepts of modern cosmology within the framework of special relativity, at a level that  is suitable for an undergraduate or high school setting. The model includes cosmic expansion governed by a universal scale factor, the Hubble relation, proper distances between galaxies and the associated recession speeds, comoving coordinates, angular and luminosity distances, and the relation between the cosmic scale factor and the cosmological redshift. It allows for a pedagogical discussion of the role of superluminal recession speeds of galaxies, the applicability of the special-relativistic Doppler formula to describe the cosmological redshift, and the validity of energy conservation for the redshifted photons. The model can be viewed as a toy version of the Milne universe.
\end{abstract}


\vfill

\bgroup
{\footnotesize
This is the version of the article before peer review or editing, as submitted by the author to the {\em European Journal of Physics}. IOP Publishing Ltd is not responsible for any errors or omissions in this version of the manuscript or any version derived from it. The Version of Record is available online at \href{https://doi.org/10.1088/1361-6404/aaf2f7}{https://doi.org/10.1088/1361-6404/aaf2f7}.}
\egroup

\maketitle 

\section{Introduction}
How is the cosmic redshift compatible with energy conservation? And why is there no contradiction between faster-than-light recession speeds of distant galaxies and the speed limit imposed by relativity? These questions are likely to arise when students attempt to reconcile newly acquired knowledge about cosmic expansion with previous knowledge about conserved quantities and the role of the speed of light in special relativity. They have also played a significant role in a recent debate in the astronomy and astronomy education communities concerning interpretations of cosmic expansion, and the best ways of teaching about these and other common questions about basic cosmology; cf.\ \cite{DavisLineweaver2004,BunnHogg2009,FrancisEtAl2007,Chodorowski2011,AbramowiczEtAl2009} and references therein. 

Most of the debate relies on advanced tools, which are not usually available in an undergraduate setting, or to high school teachers attempting to answer questions from their students. The present article describes a toy model that can be understood using no more than the standard, two-dimensional form of the Lorentz transformations of special relativity, yet reproduces, in simplified form, key properties of the cosmological standard model based on the Friedmann-Lema\^{\i}tre-Robertson-Walker spacetimes of general relativity (``FLRW cosmology,'' for short), including the concepts of cosmic time, comoving coordinates, and a cosmic scale factor. In particular, the toy model provides a useful setting for discussing superluminal recession speeds and the apparent energy loss suffered by photons travelling between galaxies. In the same way as the Milne model, which has been used as a pedagogical tool for teaching cosmology \cite{Ellis2000,Rindler2001,Mukhanov2005,GronHervik2007,Rebhan2012a}, the toy model can be used to elucidate the kinematical aspects of expanding universe model. The gravitational interactions that cause deviations from linear expansion are, naturally, beyond the scope of a model that operates within the framework of special relativity. Even so, as in other areas of physics, a basic understanding of kinematics can serve to achieve a better understanding of the full description that includes dynamics.

This makes the toy model a useful teaching tool in situations where a course does not aim at introducing the concept of the metric and the metric-based description of FLRW spacetimes or the Milne universe, but where the (undergraduate or high school) students have existing knowledge of basic special relativity. Under those circumstances, the toy model allows students to resolve apparent contradictions (the faster-than-light speeds, and apparent non-conservation of photon energies) within the framework of special relativity, using arguments that are analogous to the way the situation is described in the full FLRW formalism. We have found the toy model to be particularly helpful for high school teachers who, while not fully integrating the model into their classwork, have used it to gain a basic understanding of the physical principles involved, and to lay a foundation for simple explanations to pass on to their students. 

The general description of the toy model and its associated concepts should be accessible to anyone familiar with the basics of special relativity. The descriptions of how the model fits into the broader framework of general relativity and FLRW cosmology, on the other hand, naturally require knowledge of the relevant ideas and concepts. Readers unfamiliar with the formalism and concepts of general relativity should concentrate on the basic description of the toy model and its properties, and may safely skip sections \ref{Comoving}, \ref{LuminosityDistance} and \ref{MilneLimit}.

\section{Stitched-together inertial systems: defining cosmic time}
\label{InertialSystems}

In FLRW cosmology, the universe is populated by idealized galaxies whose mutual distances are directly proportional to a universal cosmic scale factor. In an expanding universe, the scale factor increases over time, and distances between the idealized galaxies increase accordingly. The resulting pattern of changing distances is known as the ``Hubble flow.'' In the standard cosmological coordinate system, which is adapted to the assumed symmetries of the large-scale cosmos (homogeneity and isotropy), all galaxies in the Hubble flow are locally at rest, their spatial coordinate values constant, and for each of them, its proper time is a local realisation of a global time coordinate called ``cosmic time''. For brevity, let us call a coordinate description with these two properties a {\em unified cosmic coordinate system}.

To define the toy model, we will introduce unified cosmic coordinates, putting all galaxies on an equal footing, in a simple setting, while staying as close as possible to an elementary formulation of special relativity.

Initially, let us consider only two galaxies in the Hubble flow, one of them our own galaxy, which is our observation post for studying the rest of the universe. For simplicity, assume that the relative speed of these two galaxies is constant. Then we can easily model the situation within special relativity, assigning to each galaxy an inertial coordinate system. As is usual in elementary treatments of special relativity, we restrict our attention to the plane $y=z=0$. Let one of our galaxies represent the spatial origin of an inertial system $S$ with coordinates $t,x$, the other of an inertial system $S'$ with coordinates $t',x'$, the two systems linked by the Lorentz transformations 
\bea
\label{LorentzTPrime}
ct' &=& \gamma(\beta_v)(ct-\beta_v x)\\[0.5em]
\label{LorentzXPrime}
x' &=& \gamma(\beta_v)(x-\beta_v ct)
\eea
with the usual abbreviations
\be
\label{gammaBeta}
\beta_v \equiv v/c, \;\;\;\;\;\;\; \gamma(\beta_v)\equiv \frac{1}{\sqrt{1-\beta_v^2}},
\ee
$c$ the vacuum speed of light, and $v$ the speed of the two systems' relative motion in the common $x$ direction.

At $t=0, t'=0$, the spatial origins of the two systems coincide. In line with FLRW cosmology, we call this event the big bang. At $t\ge 0$, $t'\ge 0$, the galaxies move away from each other. Throughout the rest of the article, we restrict our description to this scenario, that is, to the future light cone of the big bang. The cosmological toy model described in the following will be defined only within this region of spacetime.

What are suitable unified cosmic coordinates for our toy model? By definition, in those 
coordinates, both of the galaxies should be at rest. But we already know that the first galaxy is at rest in the coordinates of the inertial system $S$, and the second in the coordinates of $S'$. Thus, a straightforward way of defining a cosmic coordinate system in which {\em both} galaxies are at rest is to partition spacetime, using $S$ coordinates in one domain and $S'$ coordinates (with a minor modification in the shape of a constant x shift, as it will turn out) in the other. As long as the domain described using the $S$ coordinates contains the spatial origin of $S$, and the region described using the coordinates of $S'$ contains the spatial origin of $S'$, these combined coordinates will indeed describe each of the two galaxies as being locally at rest. This construction also ensures that we meet the second criterion for a unified cosmic coordinate, since for an object at rest in an inertial system $S$, proper time is measured by $t$, and for an object at rest in $S'$, proper time is measured by $t'$.

\begin{figure}[t]
\begin{center}
\includegraphics[scale=1.0]{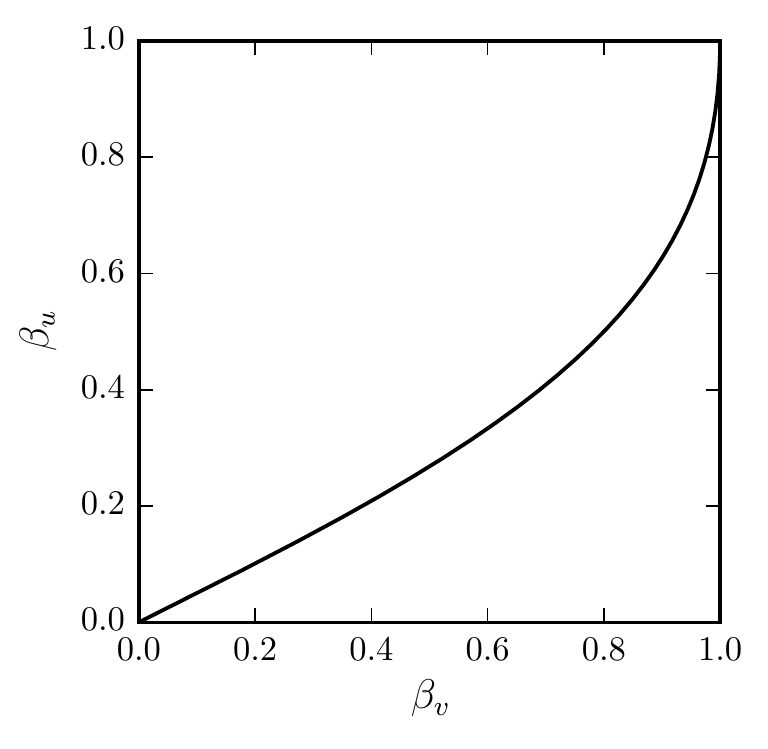}
\caption{Defining the boundary line: dependence of $\beta_u$ on $\beta_v$.}
\label{PlottingU}
\end{center}
\end{figure}

\begin{figure}[t]
\begin{center}
\includegraphics[scale=1.0]{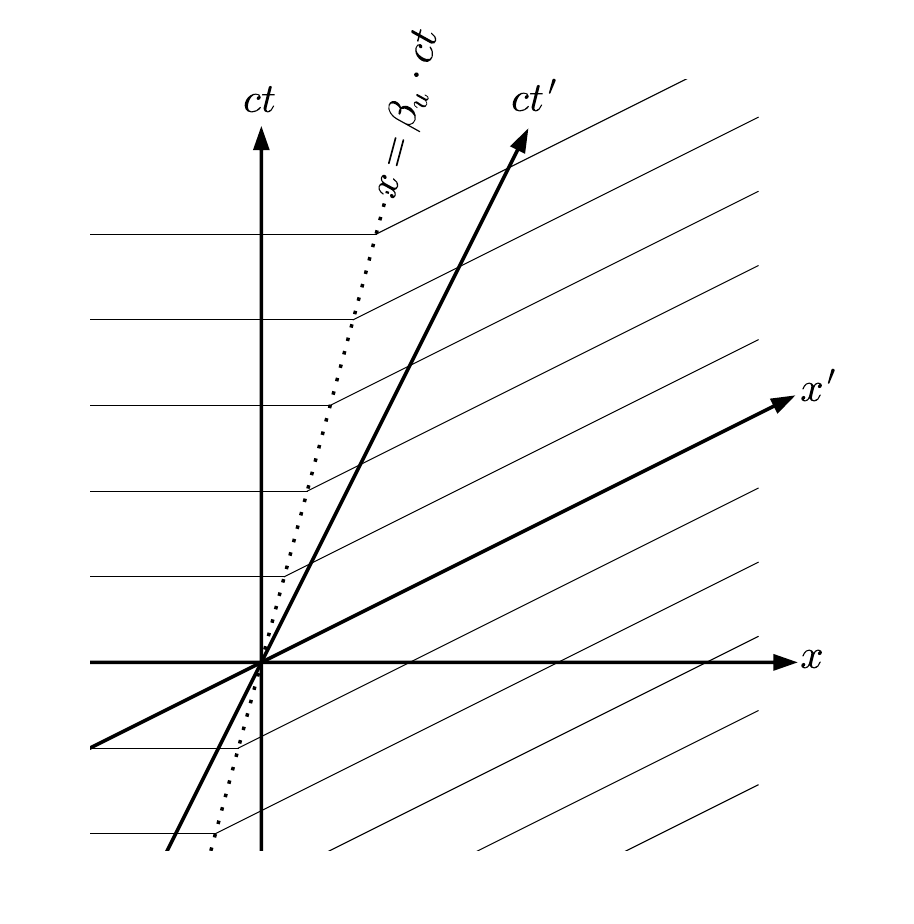}
\caption{Foliation of spacetime obtained from stitching together systems $S$ and $S'$ along $x=\beta_uct$ (dotted line), shown here for $\beta_v=0.5$.}
\label{twoStitch}
\end{center}
\end{figure}

Coordinate systems should be continuous. Spacetime coordinate values should not jump as we move from one event to a neighbouring event. Demanding continuity for the time coordinate is sufficient to define the partition we are looking for: events on the border between the $S$ and the $S'$ domain should have $t=t'$. This condition, together with the Lorentz transformation (\ref{LorentzTPrime}), yields a linear equation which is solved by the line in the  $x$-$ct$ plane,
\be
\label{stitchingLine}
x = ct\cdot \frac{\gamma(\beta_v)-1}{\gamma(\beta_v)\cdot \beta_v} \equiv ct\cdot \beta_u.
\ee
To ensure time coordinate continuity, our two separate domains will be joined along this line. Eq.\ (\ref{stitchingLine}) also
serves to define the $\beta_v$-dependent dimensionless quantity $\beta_u$. In figure \ref{PlottingU}, $\beta_u$ is plotted against $\beta_v$.
Evidently, $0\leq \beta_u \leq \beta_v$, so that the boundary line between $S$ and $S'$ is timelike, and is always located somewhere between the $t$ and the $t'$ axis. Using this result, we define our new, stitched time coordinate $\bar{t}$ via
\be
c\bar{t} = \left\{ \begin{array}{ll}
 ct& \mbox{for}\; x\le \beta_u\cdot ct\\[0.5em]
 ct' =\gamma(\beta_v)(ct-\beta_v x) & \mbox{for}\; x > \beta_u\cdot ct.
\end{array}\right.
\label{tBarStitched}
\ee
The time coordinate $\bar{t}$ defines both a global notion of simultaneity and a measure of how time passes between one moment and the next. In the technical terms of general relativity, we have defined a foliation of spacetime into spacelike surfaces and specified a lapse function. Some of the lines of constant time for our new coordinate, $\bar{t}=\mbox{const.}$, are shown in figure \ref{twoStitch} for the case $\beta_v=0.5$.
Just as we intended, our new time coordinate is a toy version of the cosmic time coordinate in FLRW cosmology: By construction, at the location of each of our two toy galaxies, $\bar{t}$ corresponds to the local time as measured by the galaxy's/the observer's own clock, with the zero point $\bar{t}=0$ corresponding to the (singular) big bang event where our galaxies' locations coincided. 

\section{Comoving coordinates}
\label{Comoving}

What the direct stitching does not automatically provide is a unified space coordinate $\bar{x}$. After all, in their respective rest frames, our galaxies are at $x=0$ and $x'=0,$ respectively, which do not coincide after the big bang, so that we cannot map both to $\bar{x}=0$. 

The simplest change that allows us to introduce a unified space coordinate is to set $\bar{x}=x$ for the region $x<0$ and $\bar{x}=x'+C$ with some constant $C$ for $x'>0$. While a spatially shifted version of the system $S'$ is not identical to $S'$, it is the next best thing, preserving key properties of $S'$ such as distances in space and time, and simultaneity. In particular, this choice ensures that both our galaxies indeed have $\bar{x}=\mbox{const.}$ -- in other words: for these two galaxies, our unified coordinate ``moves along with them,'' motivating the appellation {\em comoving coordinate}. This leaves us with the problem of how to describe the ``transition region'' between the two galaxies, whose proper distance, after all, is growing continuously. Infinitely many kinds of interpolation are possible; a simple choice is linear interpolation, 
\be
\bar{x} =  \left\{ \begin{array}{ll}
x &\mbox{for}\; x< 0\\[0.5em]
C\cdot\left(\frac{x}{\beta_vct} \right)  &\mbox{for}\;  0 \le x\le \beta_vct\\[0.5em]
x'+C = \gamma(v)(x-\beta_vct) + C  &\mbox{for}\;  x>\beta_vct.
\end{array}\right.
\ee
Using this comoving coordinate, our two galaxies are manifestly at rest in the unified coordinate system.  But the stitching comes at a price: In the transition region, coordinate distances do not correspond to physical distances any more. We do know how to determine physical distances in those regions, since we can apply the (inertial) measurements of $S$ to the left of the dividing line, and those of $S'$ to the right. But in the transition region, these measurements are decoupled from our unified spatial coordinate. Also, the comoving coordinate is not well-defined at the event $t=0,x=0$. 

In FLRW cosmology, the situation is analogous: There, all the spatial coordinates are comoving, with constant spatial coordinates characterizing the geodesics of galaxies in the Hubble flow. Physical distance measurements, notably proper distances, do not correspond to spatial coordinate distances, but must be computed using the spacetime metric. FLRW comoving coordinates, too, become ill defined at the big bang singularity, as all galaxies with finite comoving coordinate values end up at the same singular spacetime event at cosmic time zero.

As we shall see, we will never make use of the explicit form of the unified spatial coordinate in the following, as we compute proper distances, recession velocities and cosmic redshifts. Instead, for all physical distance measurements, we will fall back on the underlying inertial system. Thus, while we have included comoving coordinates here for completeness, they may be omitted in the more basic versions of teaching the toy model.

\section{Simultaneity and proper distance}
\label{SimultaneityDistance}

In order to calculate the distance between our two galaxies at some specific cosmic time, recall how distances are measured in the usual Cartesian inertial reference frames of special relativity. There, each inertial observer can use their time coordinate $t$, and thus their notion of simultaneity, to define a snapshot of space at a given moment of constant time, $t=\mbox{const.}$, as the set of all events for which the time coordinate has one specific value. Length measurements by such an observer in special relativity are tied directly to this notion of simultaneity: The length of a one-dimensional object is the coordinate distance between the object's two end points, whose positions are evaluated simultaneously. The simplest example is the derivation of relativistic length contraction, which crucially depends on the different notions of simultaneity of different inertial observers. Space as defined by simultaneity is relative, and only spacetime is observer-independent. 

\begin{figure}[t]
\begin{center}
\includegraphics[scale=1.0]{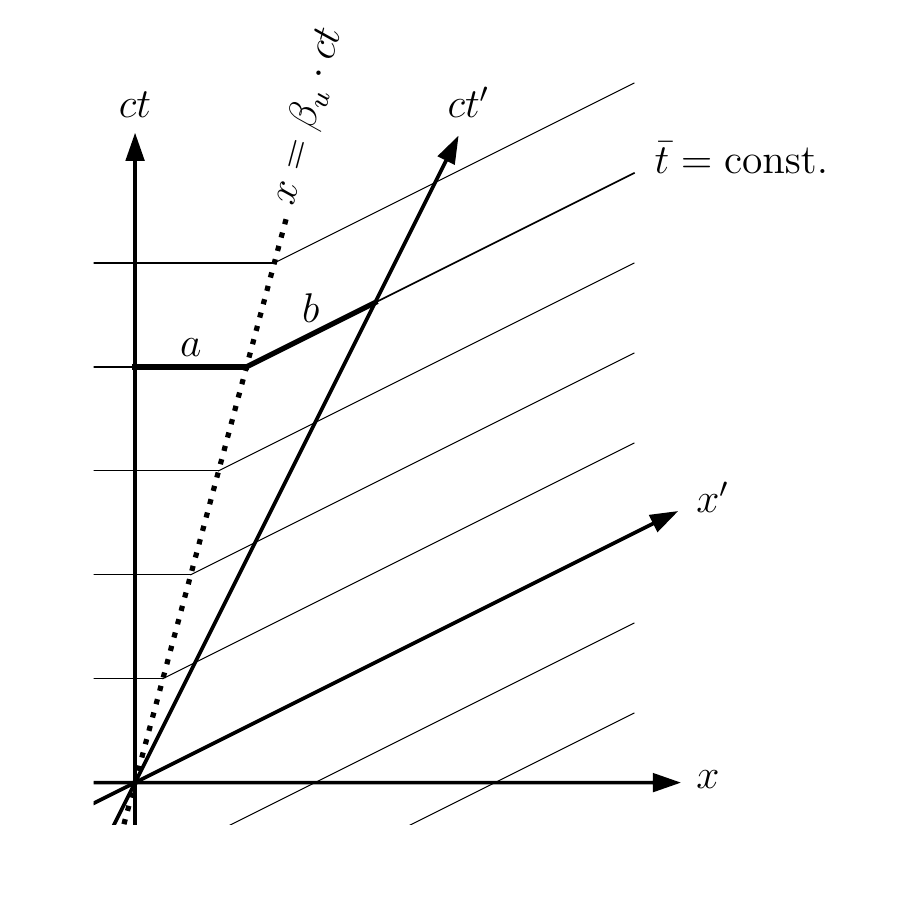}
\caption{Measuring the distance between the two galaxies at fixed time $\bar{t}=\mbox{const.}$}
\label{SnapshotSpace}
\end{center}
\end{figure}

The procedure is readily generalized to our stitched coordinate system with the setup shown in figure \ref{SnapshotSpace}. 
Our continuity requirement for $\bar{t}$ means that the snapshot $\bar{t}=\mbox{const.}$ of space at that time, defined as the set of all events that share some fixed value for $\bar{t}$, is well defined. The world lines of our two galaxies are the $t$ and the $t'$ axis, respectively. The solid line marked $\bar{t}=\mbox{const.}$ represents a snapshot of space at some particular fixed time. Since we confine our attention to the x-t-plane, that snapshot is a one-dimensional curve. The section of space separating the positions of the two galaxies is marked by the two thick line segments with lengths $a$ and $b$. 

Using simultaneity as defined by the time coordinate $\bar{t}$, the distance between the two galaxies at that particular moment is $a+b$. But for each of the two segments separately, we know how to calculate their proper length: The segment marked $a$ is simultaneous in $S$, and the segment marked $b$ is simultaneous in $S'$. Furthermore, both inertial systems are defined in the standard way, with Cartesian coordinates: x coordinate differences correspond directly to physical lengths as measured in the respective frame. This reduces our problem of determining the distance between the two galaxies to
solving linear equations for the intersections of the straight lines that make up $\bar{t}=\mbox{const.}$ with the stitching boundary (\ref{stitchingLine}). The line segment $a$ is at $\bar{t}=t=\mbox{const.}$ and thus intersects $x=\beta_u\cdot ct$ at the point $x=\beta_u\cdot c\bar{t}$. Since the starting point of $a$ is $x=0$, its length is $\beta_u\cdot c\bar{t}$. Segment $b$ is part of the line $\bar{t}=t'=\mbox{const.}$, which intersects $x=\beta_u\cdot ct$ at $t=\bar{t}/[\gamma(\beta_v)(1-\beta_v\beta_u)]$, corresponding to 
$$
x' = \frac{(\beta_u-\beta_v)}{(1-\beta_u\beta_v)}\cdot c\bar{t} = -\beta_u\cdot c\bar{t}.
$$
Since the starting point of segment $b$ is $x'=0$, its rest length, as measured in the $S'$ system, is $\beta_u\cdot c\bar{t}$, which is the same as that of $a$ measured in $S$. Thus, the result is 
\be
\label{ProperDistance}
D(\bar{t}) =a+b = 2\, \beta_u\cdot c\bar{t},
\ee
evaluated at whatever constant value for $\bar{t}$ we are considering. This distance, measured by combining measurements in two local inertial systems (one for each of our two galaxies), is a simplified version of what cosmologists call the {\em proper distance} between galaxies in the Hubble flow: distance as measured by the spacetime metric along a hypersurface of constant cosmic time. 

In analogy with FLRW cosmology, let us factor the proper distance (\ref{ProperDistance}) into a galaxy-specific part $2\,\beta_u$ on the one hand and a time-dependent part $c\bar{t}$ on the other. The latter is customarily called the {\em (cosmic) scale factor}. (Our choice to include the factor 2 not in the scale factor, but in the galaxy-specific part, is ad hoc. The scale factor can be rescaled by a constant factor without altering the physics. Observable predictions follow only from the ratio of scale factor values taken at different times.) 

The proper distance defined here in terms of the unified time coordinate is not a quantity that can be measured directly by a local observer on Earth; the same is true for the proper distance as a function of cosmic time in FLRW cosmology. We will introduce two types of distance that can be determined from observation below in section \ref{LuminosityDistance}.

\section{Recession speed and relative speed}

We define the {\em recession speed} $V$ of each of our galaxies relative to the other as the derivative of $D(\bar{t})$ with respect to $\bar{t}$, which is
\be
V\equiv \frac{\Dd D}{\Dd\bar{t}} = 2 \beta_u c.
\label{RecessionSpeed}
\ee
In particular, this means the recession speed will be superluminal for any values $\beta_v>4/5$. In the limit  $\beta_v\to 1$,  we have $V\to 2c$. In FLRW cosmology, the recession speed defined as the cosmic-time derivative of a galaxy's proper distance will also be superluminal for galaxies beyond a certain distance. 

In our toy model, it is evident that these superluminal recession speeds do not contradict the speed limit of special relativity. In all inertial frames, notably in $S$ and in $S'$, the world lines of both galaxies are timelike curves; the two galaxies move slower than the speed of light in all inertial frames. The superluminality of the recession speed (\ref{RecessionSpeed}) does not correspond to the usual special-relativistic notion of superluminal motion, that is, to a world-line with at least one space-like portion. Instead, it is a direct consequence of introducing non-inertial coordinates of the type we have introduced as unified cosmic coordinates. In particular, the recession speed (\ref{RecessionSpeed}) does not correspond to the relative speed between the two galaxies in question which, in special relativity, in the case of two objects moving uniformly within an inertial frame, can be obtained by the simple expedient of going to any inertial rest frame.

When students worry about whether or not superluminal recession speeds in an expanding universe contradict relativity, the uneasiness stems from their familiarity with the special-relativistic dictum that no object can move faster than light. (After all, the more restricted general-relativistic version that, locally, no object can directly overtake light, is not directly applicable to the situation at hand.) In the toy model, the apparent conflict is resolved by showing that, even in a situation fully within the framework of special relativity, the introduction of a unified time leads to superluminal speeds. But these speeds do not correspond to superluminal motion within any inertial system, and thus do not break the special-relativistic speed limit.

For the rest of the section, let us move beyond the metric-less toy model in order to see that this resolution is analogous to the way the apparent contradiction is resolved when treating FLRW spacetimes in the full framework of general relativity; readers unfamiliar with general relativity may want to skip ahead to section \ref{Redshifts}. Recall that in special relativity, the relative speed $v_{rel}$ of two objects, evaluated at an event ${\cal E}_1$ on the worldline of the first object and an event ${\cal E}_2$ on the worldline of the second, can be defined as follows \cite{Synge1966}: For $U_1$ the four-velocity of object 1 at the event ${\cal E}_1$, and $U_2$ that of object 2 at event ${\cal E}_2$, 
\be
\label{RelativeSpeedProduct}
 \gamma(v_{rel}/c) =  \frac{1}{c^2}\cdot\eta(U_1,U_2) \equiv \frac{1}{c^2}\cdot\left[ U_1^0U_2^0 - U_1^1U_2^1 - U_1^2U_2^2 - U_1^3U_2^3\right],
 \ee
 where we use the ``mostly minus'' convention for the metric, in which timelike four-vectors have positive norm. This definition of relative speed can be extended to general relativity, 
\be
\label{RelativeSpeedGR}
 \gamma(v_{rel}/c) =  \frac{1}{c^2}\cdot g(U_1,U_2),
 \ee
using the general spacetime metric $g$ in lieu of the Minkowski metric $\eta$ to define the scalar product. There is, however, an added difficulty: In the standard, Cartesian inertial coordinate systems special relativity, we can compare four-vectors by simply comparing their components. In curved spacetime in general relativity, four-vectors defined at different events are defined in mathematically completely distinct spaces. In order to be able to compare such different four-vectors, additional structure is needed, which allows for ''transporting'' vectors from one event to the next. In particular, this structure, known mathematically as a {\em connection}, is necessary for calculating rates of change for such vectors, given that derivatives of vectors involve the comparison of two vectors defined at (infinitesimally) nearby points. The resulting prescription for transporting vectors is called {\em parallel transport}. The result of such a transport depends on the chosen spacetime path linking the initial event and the target event if (and only if) spacetime is curved. Parallel transport along the straightest possible line (in the parlance of differential geometry, the {\em geodesic}) joining two events provides a more restricted prescription for comparing four-velocities, and thus for determining the corresponding relative speed \cite{Lanczos1923,Synge1966,Bolos2007}, which is unique for suitably simple spacetimes (in technical terms: for spacetimes without conjugate points). In particular, this definition of relative speed is unique in open FLRW universes, but non-unique in closed FLRW universes \cite{Rosquist1982}.

It can be shown using general geometric considerations that relative speeds defined in this way are always subluminal for moving objects. (Specifically, using the Lorentzian version of the Cauchy-Schwarz identity, as well as the fact that four-velocities have $g(U_i,U_i)=c^2$, it can be shown that the RHS of (\ref{RelativeSpeedGR}) is always greater than 1, so the equation indeed has a real solution $v_{rel} < c$.)

Thus, the apparent contradiction --- superluminal speeds for moving objects? --- is resolved in the same way in FLRW as it is in our toy model. In both cases, the recession speed (\ref{RecessionSpeed}) is not a valid generalization of the relative speed of special relativity. In fact, the toy model shows that it does not correspond to the relative speed even in situations that are defined in a flat universe governed by special relativity. Thus, there is no requirement for recession speeds of the type (\ref{RecessionSpeed}) to obey the relativistic speed limit imposed on motion relative to an inertial observer.

\section{Cosmic redshift}
\label{Redshifts}
Light emitted from a galaxy in the Hubble flow, and received by an observer in another such galaxy, is redshifted in a systematic way. In our toy model, we can evaluate this {\em cosmological redshift} in either one of the inertial systems (since each is defined globally, after all). It follows directly that the redshift 
$$
z \equiv \frac{\lambda_0-\lambda_e}{\lambda_e},
$$
is described by the longitudinal Doppler formula of special relativity,
\be
z+1=\sqrt{\frac{1+\beta_v}{1-\beta_v}}.
\label{SRDoppler}
\ee
Here, $\lambda_e$ is the wavelength of light as it is emitted by one of the galaxies, as measured in the emitting galaxy's rest frame, and $\lambda_0$ the wavelength at which the same light is received as it arrives at the other galaxy, again as measured in that galaxy's rest frame.

This cosmological redshift can be expressed in an alternative way as follows. Assume that a light signal leaves the second galaxy, which is located at the spatial origin of $S'$, at time $\bar{t}_e=t'_e,$ and travels toward the first galaxy, that is to the spatial origin of $S$. By the Lorentz transformation (\ref{LorentzTPrime}), the emission time is $t_e =t'_e\cdot \gamma(\beta_v)$ in the coordinate time of system $S$. From the location $x_e=\beta_v\cdot ct_e$ of the second galaxy in system $S$, it will take the time interval $x_e/c$ for the light to reach the spatial origin of $S$. Thus, the arrival time is
\be
\bar{t}_0 = t_0 = t_e + \frac{x_e}{c} = (1+\beta_v)\gamma(\beta_v)\cdot \bar{t}_e = \sqrt{\frac{1+\beta_v}{1-\beta_v}}\cdot \bar{t}_e.
\ee
The square root on the right is exactly the Doppler shift (\ref{SRDoppler}), while the time coordinate values are proportional to the cosmic scale factor, $a(\bar{t})=c\bar{t}$. Evidently, the cosmological redshift can be expressed as
\be
1 + z = \frac{\bar{t}_0}{\bar{t}_e} = \frac{a(\bar{t}_0)}{a(\bar{t}_e)}.
\label{ScaleFactorRedshift}
\ee
The ratio between the received and emitted wavelength is the same as the ratio of the scale factors at the time of reception and the time of emission. 

For those readers who are familiar with the general-relativistic formalism for describing corresponding cosmological models, it is worth noting how these features of our toy model correspond to more general properties of cosmological models. In FLRW cosmology, the scale factor values at the emission and reception time and the cosmic redshift are linked as in equation (\ref{ScaleFactorRedshift}), although the scale factor will not, in general, be directly proportional to cosmic time. The same cosmological redshift can also be expressed using the generalized relative velocity $v_{rel}$, obtained by parallel transport of the emitting galaxy's four-velocity along the light-like geodesic linking the emission and the reception event, and the special-relativistic Doppler formula (\ref{SRDoppler}) \cite{Narlikar1994,Liebscher2007,BunnHogg2009,CookBurns2009,Kaya2011}.

What about the apparent loss of energy by photons travelling in an expanding universe? The toy model reduces this to the familiar situation of two observers in relative motion. This is obvious if we put aside the cosmological context and our unified description, and revert to the usual view of two observers in inertial frames $S$ and $S'$. 
Regarding $S$ and $S'$ as properly separate, no student who has understood the Doppler effect should be surprised by the fact that an observer in an inertial frame $S$ measures one value for the energy (equivalently: for the wavelength) of an emitted photon, while an observer in an inertial frame $S'$ measures a lower energy. The different energy values are the consequence of different inertial frames of reference, not of any physical process changing the photon. The fact that we can stitch those two inertial systems together using a unified cosmic coordinate system which hides that relative motion, but retains its consequences for measurements of photon energy, is a poignant example of why we should not expect classical energy conservation to hold in the general, non-inertial coordinate systems allowed in general relativity, and used in FLRW cosmology. 

\section{Angular distance and luminosity distance}
\label{LuminosityDistance}

The distances introduced so far --- comoving coordinate distance in section \ref{Comoving} and proper distance in \ref{SimultaneityDistance} --- cannot be measured directly using only local observations. Just like their counterparts in FLRW cosmology, they are useful in formulating the underlying model, but additional definitions are needed to forge the link with observations.

One definition of an observable distance is that of {\em angular distance} $d_{ang}$. If we are observing a one-dimensional object that, transverse to the direction of observation, has the length $\Delta L$, and if the angular size we are observing for that object (in radians) is $\Delta\phi$, then the angular distance of that object is given by the equation 
\be
\Delta L = d_{ang}\cdot \Delta\phi.
\ee
Since lengths that are transverse to the direction of motion remain unchanged in the transition from one inertial system to another, $d_{ang}$ is the same as the distance as measured in our inertial system $S$ at the emission time $t_e$, using the standard coordinates defined in special relativity. For light that reaches the observer in the origin of $S$ at local cosmic time $\bar{t}=t$, that emission time is $t_e = \bar{t}\cdot (1+\beta_v)^{-1}$, so
\be
d_{ang}(\bar{t}) = c\bar{t}\cdot\frac{\beta_v}{1+\beta_v} = \frac{\beta_v/\beta_u}{2(1+\beta_v)}\cdot D(\bar{t}),
\ee
where the rightmost expression relates the angular distance and the proper distance $D(\bar{t})$ defined in (\ref{ProperDistance}).

Another useful cosmological distance measure is the {\em luminosity} distance. Imagine that an object at rest in the receding system $S'$ is emitting light in an isotropic fashion, and that an observer at rest relative to that objects measures a total luminosity of $L$ in terms of energy emitted per unit time. How bright will that object appear to an observer at rest at the origin of $S$, concretely: what energy flux $F$ from that object will such an observer measure?

Recall that for astronomical observations, the energy flux is the energy per unit time per collecting area of a telescope pointed directly at the object in question. If the object were at rest in $S$, the energy $F\cdot\Delta A$ collected by our telescope with collecting area $\Delta A$ per unit time would be the luminosity $L$, rescaled by the ratio of the collecting area $\Delta A$ and the total spherical area $4\pi d_{ang}^2$ over which the radiation is spread out once it has travelled the distance $d_{ang}$. The result would be the usual inverse square law
\be
F = \frac{L}{4\pi d_{ang}^2}.
\ee
But since the object is moving, there are three additional special-relativistic effects, cf. section 4.4 in Rindler's book \cite{Rindler2001}, which can best be understood by looking at the radiation as a stream of photons. As derived in section \ref{Redshifts}, the cosmic redshift reduces the energy of each single photon by a factor $(1+z)^{-1}$. In addition, the rate at which photons are emitted as measured by an observer at rest in $S'$ and the rate at which photons reach an observer at rest in $S$ are related by the same factor; this is readily visualized by imagining a steady stream of equidistant photons, where the inter-photon distance measured from $S$ and from $S'$ are related in the same way as photon wavelengths. Finally, there is relativistic aberration: A photon emitted within the system $S'$ at an angle of $\theta'$ to the $x'$ axis will be seen as moving at an angle $\theta$ to the $x$ axis by an observer in $S$, where
\be
\tan(\theta/2) = \sqrt{\frac{1-\beta_v}{1+\beta_v}}\cdot\tan(\theta'/2).
\ee
By differentiating this formula, it follows that small angular differences $\Dd\theta'$ and $\Dd\theta$ measured by the two observers are related by
\be
\Dd\theta = \frac{\Dd\theta'}{\gamma(\beta_v)(1+\beta_v\:\cos\theta')}.
\ee
By that argument, radiation emitted into a small solid angle $(\Dd\theta')^2$ as judged by an observer in $S'$ will be measured as emitted into small solid angle $(\Dd\theta)^2$ by an observer in $S$. For the case we are interested in here, that is, radiation emitted by the galaxy in $S'$ backwards, towards us, as it moves away from us, at $\theta'=\pi$, we have
\be
(\Dd\theta)^2 = \frac{1+\beta_v}{1-\beta_v}\cdot(\Dd\theta')^2.
\ee
Defining our telescope collecting area by $\Delta A = \pi (d_{ang}\cdot\Dd\theta)^2$, our telescope receives light emitted, as described in $S'$, into the solid angle
\be
\pi\cdot(\Dd\theta')^2 = \left(\frac{1-\beta_v}{1+\beta_v}\right)\frac{\Delta A}{d_{ang}^{\,2}}.
\ee
Since, in $S'$, light emission is isotropic, the fraction
\be
\frac{\pi\cdot(\Dd\theta')^2}{4\pi} = \left(\frac{1-\beta_v}{1+\beta_v}\right)\frac{\Delta A}{4\pi d_{ang}^{\,2}}
\ee
of the energy flux emitted over the complete sphere $4\pi$ reaches our telescope, so this aberration effect on its own would modify the inverse square law to yield
\be
F = \left(\frac{1-\beta_v}{1+\beta_v}\right)\frac{L}{4\pi d_{ang}^{\,2}}.
\ee
Combining all three relativistic effects, we find that
\be
F =  \left(\frac{1-\beta_v}{1+\beta_v}\right)^2\frac{L}{4\pi d_{ang}^{\,2}} = \frac{L}{4\pi d_{ang}^{\,2} (1+z)^4},
\ee
which is the same relation between luminosity, energy, flux and angular distance as in FLRW cosmologies. It is customary to define the luminosity distance
\be
d_{L} = d_{ang}^{\,2} (1+z)^2,
\ee
for which the inverse square law takes on the simpler form
\be
F = \frac{L}{4\pi d_{L}^{\,2}}.
\ee

\section{More than two galaxies}
\label{MoreThanTwo}

At least for galaxies lined up along the x axis, our calculations are readily extended from two galaxies to an arbitrary number of them. (The same reasoning can readily be applied to radial separation in a spherically-symmetric situation, but that is beyond the scope of this article.) Denote these galaxies by $G^i$, for $i=0,1,2,\ldots,$ and let them move along the x direction at constant speeds $v_i$, with the ordering $v_0=0<v_1<v_2< v_3 < \ldots$ and corresponding expressions $\beta_i\equiv v_i/c$, $\gamma_i\equiv \gamma(\beta_i)$. Introduce inertial systems $S^{(i)}$, with $S\equiv S^{(0)}$ the one in which we, the observers, are at rest. Let each galaxy $G^i$ be at rest in the spatial origin of $S^{(i)}$. All our galaxies should be at the same location at $t=0,x=0$, our version of the big bang. Let the coordinates of the inertial system $S^{(i)}$ for the x-t-plane be $t^{(i)}$ and $x^{(i)}$. Lorentz transformations between the $S^{(i)}$ and $S$ have the simple form (\ref{LorentzTPrime})--(\ref{LorentzXPrime}), with $\beta_i,t^{(i)}, x^{(i)}$ substituted for $\beta_v,x',t'$.

Our recipe for defining unified cosmic coordinates is readily generalized: In the neighbourhood of $G^i$, we choose the definition of simultaneity associated with the inertial system $S^{(i)}$, and we stitch the different notions of simultaneity together to define our new unified cosmic time coordinate. Just as in the two-galaxy case, in a region around $G^i$, a newly unified cosmic time coordinate $\bar{t}$ will have the same value as the time coordinate $t^{(i)}$ of $S^{(i)}$ in some region around $G^i$. 

\begin{figure}[htbp]
\begin{center}
\includegraphics{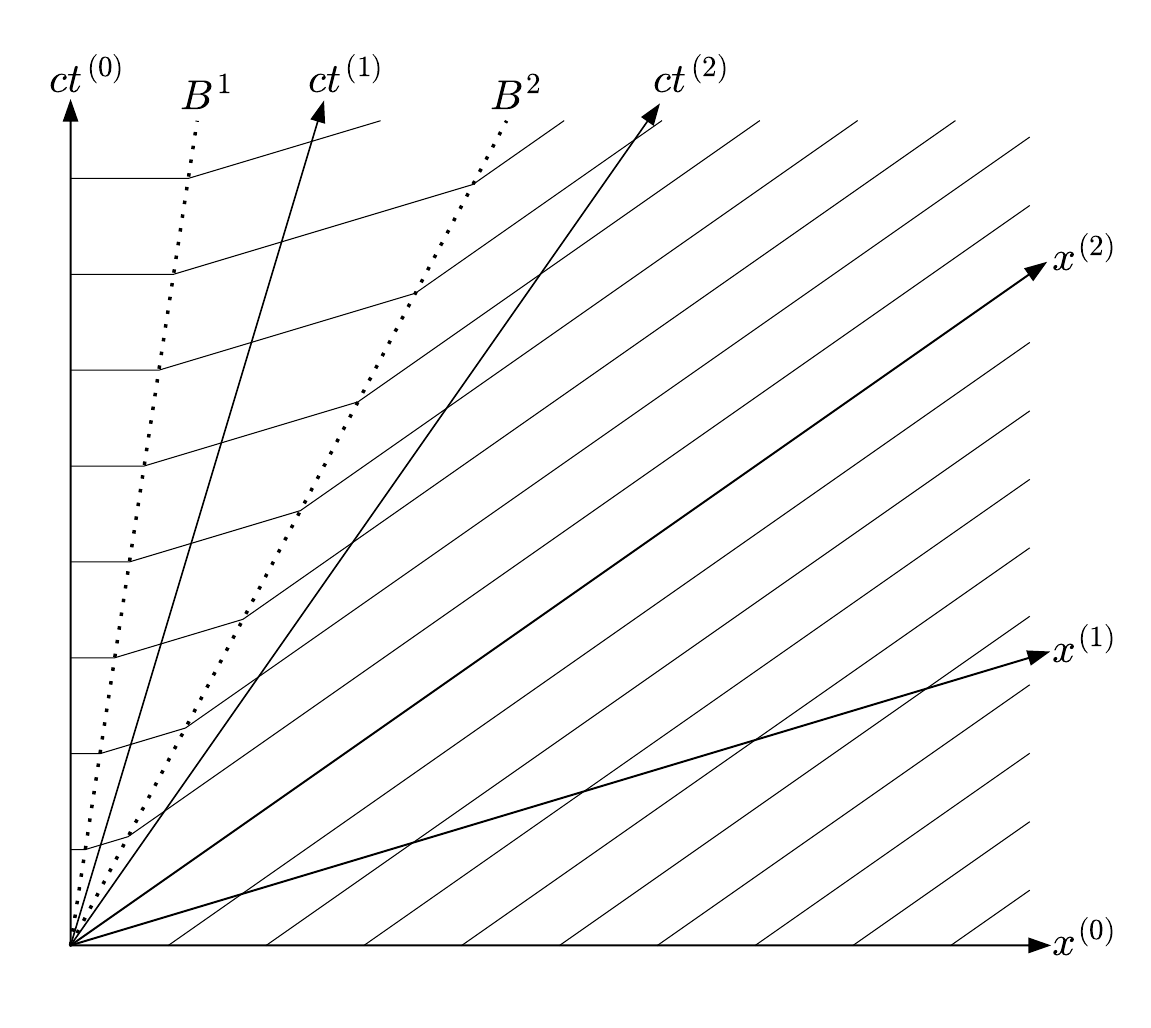}
\caption{Three galaxies and their associated inertial systems with spatial axes $x^i$ and time axes $ct^i$, separated by boundary lines $B^i$. Also shown are some lines of constant unified time coordinate $\bar{t}$. }
\label{ThreeSystems}
\end{center}
\end{figure}

The boundaries between the various regions of local time are again determined by demanding continuity for the cosmic time coordinate $\bar{t}$. From the Lorentz transformation (\ref{LorentzTPrime}), we can see that a line of constant $t^{(i)}=\bar{t}$ in the x-t-plane satisfies the equation
\be
x = \frac{c(t-\bar{t}/\gamma_i)}{\beta_i}
\label{iconstTLine}
\ee
when expressed in the coordinates of $S$. For systems $S^{(i)}$ and $S^{(i+1)}$ the corresponding lines, for the same $\bar{t}$, intersect at
\be
t = \bar{t}\cdot\frac{\gamma_{i+1}\beta_{i+i}-\gamma_i\beta_i}{\gamma_{i+1}\gamma_i(\beta_{i+1}-\beta_i)}.
\ee
We can use this together with (\ref{iconstTLine}) to eliminate $\bar{t}$ and construct the line along which the systems $S^{(i)}$ and $S^{(i+1)}$ must be stitched together to ensure continuity of the unified time coordinate, namely
\be
x = ct\cdot \frac{\gamma_{i+1}-\gamma_i}{\beta_{i+1}\gamma_{i+1} - \beta_i\gamma_i} \equiv ct\cdot \beta_{ui}.
\label{StitchLineI}
\ee 
This is the generalization of the boundary line which, for the case of two galaxies, was given by (\ref{stitchingLine}). Let us call the boundary line defined by (\ref{StitchLineI}) $B^{i+1}$. For the special case of three linked systems, the axes, boundary lines, and lines of constant unified time coordinate $\bar{t}$ are shown in figure \ref{ThreeSystems}.

Next, we calculate the proper distance between $G^i$ and $G^{i+1}$ in the same manner as in section \ref{SimultaneityDistance}, namely along the segmented line $\bar{t}=\mbox{const.}$ in the region between $G^i$ and $G^{i+1}$. The first segment is simultaneous as defined within the system $S^{(i)}$, the second segment is simultaneous in $S^{(i+1)}$. The $S$-coordinates of $G^{i}$ at unified cosmic time $t^{(i)}=\bar{t}$ are
\be
t=\gamma_i\,\bar{t} \;\;\;\mbox{and}\;\;\; x = \beta_i\,\gamma_i\cdot c\,\bar{t}.
\label{GiWorldLine}
\ee
The line segment that is simultaneous at $t^{(i)}=\bar{t}$, whose equation in $S$-coordinate was given in eq. (\ref{iconstTLine}), intersects the boundary line at 
\be
t=\bar{t}\cdot\frac{\beta_{i+1}\gamma_{i+1}-\beta_i\gamma_i}{\gamma_{i+1}\gamma_i(\beta_{i+1}-\beta_i)}, \;\;\;
x=c\bar{t}\frac{\gamma_{i+1}-\gamma_i}{\gamma_i\gamma_{i+1}(\beta_{i+1}-\beta_i)}.
\ee
Using the simple Lorentz transformations, this intersection has the x coordinate value
\be
x^{(i)}=\frac{[\gamma_{i+1}\gamma_i(1-\beta_i\beta_{i+1})-1]}{\gamma_{i+1}\gamma_i(\beta_{i+1}-\beta_i)}\cdot c\bar{t} \equiv \bar{\beta}_{ui}\cdot c\bar{t}
\label{xiValue}
\ee
in the system $S^{(i)}$; since galaxy $G^{i}$ is at rest at $x^{(i)}$, the right-hand side of (\ref{xiValue}) is also the length of the line segment joining the intersection point and $G^i$, measured in $S^{(i)}$. An analogous calculation shows that the line segment defined by the intersections of $t^{(i+1)}=\bar{t}$ with the line (\ref{StitchLineI}) and with $G^{i+1}$ has the same proper length. 
Thus, the proper distance between the galaxies $G^{i}$ and $G^{i+1}$, evaluated at some time $\bar{t}$, is
\be
D_{i,i+1}(\bar{t}) = 2\,\bar{\beta}_{ui}\cdot c\bar{t}.
\label{IncrementalStitched}
\ee
The proper distance between our own galaxy at the origin of $S$ and the galaxy $G^i$ is obtained by 
summing up the proper lengths of all the line segments along the way, which yields
\be
D_{0,i}(\bar{t}) = 2\left(\sum_{j=0}^{i-1}\bar{\beta}_{uj} \right)\cdot c\bar{t}.
\label{ProperDistancesI}
\ee
The recession speed of the galaxy $G^i$ from the origin of $S$ is 
\be
V_i \equiv \frac{\Dd D_{0,i}}{\Dd\bar{t}}.
\ee
But since, by (\ref{ProperDistancesI}), $D_{0,i}$ is directly proportional to $\bar{t}$, all the recession speeds $V_i$ for galaxies in the Hubble flow obey the relation
\be
V_i  = H\cdot D_{0,i}.
\label{HubbleRelation}
\ee
with a galaxy-independent parameter
\be
H(\bar{t}) = {1}/{\bar{t}}.
\label{HubbleParameter}
\ee
Relation (\ref{HubbleRelation}) is the Hubble relation, which is at the core of FLRW cosmology, and led to the discovery of cosmic expansion in the first place. The specific form of the Hubble parameter $H(\bar{t})$ in equation (\ref{HubbleParameter}) is that for linear expansion. For a realistic, matter-filled universe, the functional form of $H(\bar{t})$ will be different; in some interesting cases, though, such as a simple radiation-filled universe, or the Einstein--de-Sitter matter-filled universe, the difference is only a small constant factor. 

In the linear case we have obtained here, the Hubble parameter evaluated at any fixed time $\bar{t}$ is the inverse of the age of the universe $\bar{t}$ at that time. In the more general case, the inverse value of the Hubble parameter will be still be linked to the age of the universe, and provide a typical time scale for the evolution of the cosmos as a whole.

For each galaxy pair $G^i$ and $G^j$, the proper distance
\be
D_{i,j}(\bar{t}) = 2\left(\sum_{k=i}^{j-1}\bar{\beta}_{uk} \right)\cdot c\bar{t}\ee
is proportional to the same cosmic scale factor,
\be
a(\bar{t}) = c \bar{t}, 
\ee
which is defined only up to a constant factor. In FLRW cosmology, all pairwise proper distances of galaxies in the Hubble flow depend on a universal scale factor $a(\bar{t})$. In general, that scale factor is a nonlinear function of cosmic time $\bar{t}$. 

Our calculations of the relationship between the redshift, arrival and emission time of light signals in section \ref{Redshifts}, and of angular and luminosity distances in terms of cosmic time in section \ref{LuminosityDistance}, did not depend on the unified coordinates linking the two galaxies involved, but only on their description within the global inertial frame $S$. Hence, the results we obtained are valid for all the galaxies $G^i$. In particular, for all these galaxies, the cosmological redshift is determined by the cosmic scale factor ratio at time of emission and time of reception on the one hand, and by the relativistic relative speed of the galaxies and the Doppler formula (\ref{SRDoppler}) on the other.

The linear expansion law $a(\bar{t})=c\bar{t}$ corresponds to a matter-free universe with no gravitational dynamics. In consequence, this is not a suitable model for the long-term evolution or for the  history of our own universe. It does, however, provide a good approximation for cosmic expansion in our direct neighbourhood --- for galaxies that are sufficiently near to us that light-travel times, and hence look-back times, to be small. This corresponds to the usual approximation of substituting the Hubble constant (that is, the Hubble parameter evaluated at the present time) for the Hubble parameter in the Hubble relation (\ref{HubbleRelation}), and interpreting the radial proper distance as an ordinary Euclidean distance.

On larger scales, the toy model corresponds to a purely kinematical description of the universe, with gravity ''switched off''. As such, it is analogous to a ''kinematics first'' approach to teaching mechanics, which begins with particles moving at constant velocity before moving on to forces and dynamics. Linear expansion allows for an understanding of basic properties of cosmic expansion, such as the role of cosmic time and the associated proper distance, apparent superluminality and apparent energy loss of photons, in a simple, controlled situation, before gravity and cosmic dynamics are ''switched on'' for a more realistic, but also more complicated functional form of the scale factor $a(\bar{t})$.

\section{The Milne limit}
\label{MilneLimit}

In the limit of infinitely closely spaced galaxies, our toy model converges to the Milne universe. This model spacetime was originally, in the 1930s, advanced as an alternative to the cosmological models of general relativity. The modern view of the Milne universe is as the zero-density limit of the Friedmann-Lema\^{\i}tre-Robertson-Walker (FLRW) spacetimes that underlie the current models of an expanding universe. The Milne universe can be viewed as a chain of inertial systems moving at infinitesimal relative speeds \cite{Milne1934a}, so we can recover it from our toy model if we let all distances between neighboring galaxy pairs $i$ and $i+1$ become infinitesimal. Let $S(\beta)$ be the frame that moves at a speed $v=\beta c$ relative to $S$. We use $\beta$ to parameterize, continuously, our infinitely many galaxies. 

We choose a new time coordinate $\bar{t}$ as follows: At each event that, in the frame $S$, has coordinate values $(x,t)$, let $\bar{t}$ be the coordinate time since the big bang as measured in that particular frame $S(\beta)$ whose spatial origin intersects the event $(x,t)$, namely the frame $S(\beta=x/ct)$. A Lorentz transformation yields
\be
\bar{t} = \sqrt{t^2-(x/c)^2}.
\label{tbartx}
\ee
This defines our notion of simultaneity. The proper distance between adjacent galaxies parameterized by $\beta$ and $\beta+\Dd\beta$ is the infinitesimal limit of (\ref{IncrementalStitched}), 
\be
\Dd D = \frac{c\,\bar{t}}{1-\beta^2}\cdot\Dd\beta.
\ee
This is readily integrated from the $\beta=0$ worldline to a particle with parameter value $\beta$ to give
\be
D(\beta) = c\bar{t}\cdot \tanh^{-1} \beta,
\ee
where $\beta=0$ marks the spatial origin of our system $S$. Using $\beta=x/ct$, we can write down the complete transformations from $x,t$ to $D,\bar{t}$ coordinates and back:
\begin{eqnarray}
ct &=& c\bar{t} \cdot\cosh(D/c\bar{t})\\[0.5em]
x &=&  c\bar{t} \cdot\sinh(D/c\bar{t})
\end{eqnarray}
and, for the inverse transformations, (\ref{tbartx}) and
\be
D = c \sqrt{t^2-(x/c)^2}\cdot \tanh^{-1}(x/ct).
\ee 
Without making explicit use of the concept of the metric, we have re-derived the usual relations between private coordinates $\bar{t},D$ and public coordinates $t,x$ in the Milne universe. 

Several authors have made use of the Milne universe to clarify specific aspects, or argue for specific interpretations, of cosmic expansion \cite{Ellis2000,Rindler2001,Mukhanov2005,GronHervik2007,Rebhan2012a}. The pedagogical role played by the Milne universes in those texts is similar to the role outlined for the toy model in section \ref{MoreThanTwo} --- as a kinematical model useful for clarifying certain fundamental properties of cosmic expansion before moving on to the more realistic, matter-filled FLRW models that include dynamical effects. However, the reliance of these texts on the concept of a spacetime metric limits the Milne universe's usefulness in an introductory undergraduate setting where cosmic expansion is discussed without delving into the formalism of general relativity. This is where the toy model presented in this article is hoped to be of use. As long as your students master the Lorentz transformations, they can understand the basics as well as some of the more subtle properties of cosmic expansion.

\section{Discussion}
\label{Discussion}

The toy model described in this article reproduces (linear) scale factor expansion, the Hubble relation, the relation between the cosmic scale factor, relativistic relative speed, and the cosmological redshift, as well as comoving coordinates.

While the model is grounded in special relativity, it does make use of unusual, stitched coordinates. Does this already represent a transition from special to general relativity, or at least a departure from special relativity? No more than, say, the introduction of spherical coordinates in describing an orbiting planet requires a departure from classical mechanics. We do not ascribe any new physical significance to the new coordinates, and have referred back to the respective inertial systems whenever we measured time intervals or proper distances. In fact, as a famous historical debate about Einstein's theory and the meaning of covariance demonstrated, not even the stronger criterion of allowing for general coordinates and demanding general covariance is sufficient to transport us into the realm of general relativity \cite{Norton1993}.

Are superluminal recession speeds of the type (\ref{RecessionSpeed}) incompatible with relativity, in particular with special relativity? 
In the context of the toy model, it is clear that recession speeds, defined as the cosmic-time derivatives of proper distances, are not relative speeds in the sense of special relativity. Analogously, the recession speeds of FLRW cosmology are not the same as relativistic relative speeds, defined using parallel transport along geodesics. Thus, there is no reason why recession speeds should obey the light-speed limit. There are, after all, infinitely many coordinate-dependent entities with the physical dimension of a speed that do not and need not obey this limit.

In contrast, from the relativistic definition of relative speeds, evaluated using a global inertial system in the case of our toy model and parallel transport along geodesics in FLRW cosmology, we recover properly subluminal relative speeds, in terms of which the cosmological redshift can be expressed as a Doppler effect. In the words of Bernard Schutz \cite{Schutz2004}, the cosmological redshift is simply how the usual Doppler shift works in an expanding universe.

What about the apparent loss of energy by photons travelling in an expanding universe? The toy model reduces this to the familiar situation of two observers in relative motion. This is obvious if we put aside the cosmological context and our unified description, and revert to the usual view of two observers in inertial frames $S$ and $S'$. Students who have never heard of the possibility of a unified description would never be confused by the fact that an observer in an inertial frame $S$ measures one value for the energy (equivalently: the wavelength) of an emitted photon, while an observer in an inertial frame $S'$ measures a lower energy. The different energy values are the consequence of different frames of reference, not of any physical process involving the photon. 

The toy model provides a bridge between this familiar situation and the less familiar one in FLRW cosmology, where the usual cosmological coordinate system obscures the fact that we are dealing with seamlessly stitched-together frames that are in relative motion. In the toy model, the situation is clear: the apparent energy loss is an effect of unifying two inertial systems into one, and no less a consequence of changing reference frames than when $S$ and $S'$ are considered as completely separate. The lesson carries over to FLRW cosmology: Once more general coordinate systems are admitted, we need to be careful when considering apparent losses of photon energy. In the toy model, the apparent loss is a simple consequence of the two observer's relative motion. Energy is not conserved between coordinate systems in relative motion. In FLRW cosmology, the same argument can be made using the relativistic relative speeds of pairs of galaxies, compared via parallel transport along the photon's geodesic. Even for teachers who do not choose to use the toy model in their classes, it still provides a basis for answering student questions about superluminal speeds and the apparent energy loss of photons in an expanding universe on the basis of special relativity, secure in the knowledge that those answers can be expanded into a fully-fledged toy model if needed, and that they are indeed analogous to the explanations provided in general relativity proper. 

Taken together, the toy model presented here allows students to understand numerous important properties of FLRW cosmology, using no more than a knowledge of the basic concepts of special relativity: the mathematics and meaning of the Lorentz transformations in the x--t plane.

\section*{Acknowledgements}

I would like to thank Thomas M{\"u}ller and Richard Toellner 
for helpful discussions.

\end{document}